\documentclass[superscriptaddress,aps,prd,twocolumn,showpacs,nofootinbib]{revtex4-1}
\usepackage{etex}
\usepackage{amsmath,amssymb,amsthm}
\usepackage[colorlinks=true,citecolor=blue,urlcolor=blue]{hyperref}
\usepackage[pdftex]{graphicx}  
\usepackage{verbatim}   
\usepackage{color}      
\usepackage{subfigure}  
\usepackage{hyperref}   
\usepackage{mathrsfs}
\usepackage{booktabs}
\usepackage[font=footnotesize,labelfont=bf]{caption} 
\usepackage{wrapfig}
\begin{document}
\title{Stochastic gravitational wave background from accreting primordial black hole binaries during
early inspiral stage}
\author{Arnab Sarkar}         \email{arnabsarkar@bose.res.in, arnab.sarkar14@gmail.com}
\affiliation{Department of Astrophysics and Cosmology, 
S. N. Bose National Centre for Basic Sciences, JD Block, Sector III, 
Salt lake city, Kolkata-700106}
\author{K. Rajesh Nayak }  \email{rajesh@iiserkol.ac.in}
\affiliation{Indian Institute of Science Education and Research, Mohanpur, West Bengal-741246}
\author{A. S. Majumdar}   \email{archan@bose.res.in, archan@boson.bose.res.in}
\affiliation{Department of Astrophysics and Cosmology, 
S. N. Bose National Centre for Basic Sciences, JD Block, Sector III, 
Salt lake city, Kolkata-700106}
\date{\today}
\begin{abstract}
\begin{center}
 \textbf{Abstract}
\end{center}
\begin{small}
We investigate the stochastic gravitational wave background produced by primordial
black hole binaries during their early inspiral stage while accreting high-density radiation surrounding them in the early universe. We first show that the gravitational wave amplitude produced from a 
primordial black hole binary has correction terms because of the rapid rate of increase in masses of the primordial black holes. These correction terms arise due to non-vanishing first and second time derivatives of the masses and their contribution to the overall second time derivative of quadrupole-moment tensor. We find that some of these correction terms are not only significant in comparison with the main term but may be even dominant over the main term for certain ranges of time in the early Universe. The significance of these correction terms  persists for the overall stochastic gravitational wave background produced from them. We show that the spectral density produced from
such accreting primordial black hole binaries lie within the detectability range of some present and future gravitational wave detectors. 
\end{small}
\end{abstract}
\maketitle
\begin{small}
\section{Introduction}

The last couple of years have seen several detections of gravitational waves from binary black
hole mergers since the first report by the LIGO and VIRGO scientific collaborations \cite{Abbott_et_al, Ligo-virgo2, Ligo-virgo3, Ligo-virgo4, Ligo-virgo5, Ligo-virgo6}. Besides detection of these individual sources,  the stochastic gravitational wave backgrounds generated from unresolvable cosmological and astrophysical sources
have also aroused interest. Among various sources of cosmological stochastic gravitational wave backgrounds, primordial black hole binaries formed in the early Universe are of considerable importance. Primordial black holes (PBHs) are produced in the early Universe by direct gravitational collapse of the regions  containing sufficiently high density fluctuations of relativistic matter or radiation. It has been argued in some works that PBHs could survive up to present times and form a significant constituent of dark 
matter~\cite{Carr_et_al, Juan}.  It has also been argued \cite{Sasaki_et_al, Bird_et_al, Blinnikov_et_al, Nishikawa_et_al} that PBHs comprise the black hole merger event GW150914, leading to the first direct detection of gravitational waves.

One of the main mechanisms of formation of PBHs is the density fluctuations originating from
the quantum vacuum fluctuations during  inflation~\cite{Orlofsky_et_al}.  After the end of the inflation, the Universe 
entered a phase of decelerated expansion resulting in the density fluctuations re-entering the Hubble horizon. For a sufficiently large amplitude of fluctuation, Jeans-instability was triggered leading
to the fluctuation collapsing to a PBH  \cite{Harada_et_al, Harada}. Further, massive PBHs could also be formed due to collapse of large curvature perturbations generated during hybrid inflation~\cite{Clesse_et_al_3}. A significant fraction of PBHs could have formed binaries which emitted gravitational waves in the course of gradual shrinking and merger~\cite{Raidal2}.
 Various aspects of stochastic gravitational waves from PBHs have been studied~\cite{Nakama_et_al, Clesse_et_al_2, Wang_et_al, Mandic_et_al, Raidal, Hayasaki_et_al}. Most of these works are related to the PBHs that formed during the late Universe [e.g.-\cite{Clesse_et_al, Clesse_et_al_2}], while a few have discussed some early Universe effects ~\cite{Raidal,Yacine_et_al}.
  
There are certain key differences between the rate of formation of binaries of black holes in the early and late Universe. In the early Universe,  the rate of expansion of Universe was so rapid that it had a significant effect on the rate of PBH binary formation. Moreover, the density of the background radiation was robust leading to a considerably higher rate of accretion. It has been argued that accretion of surrounding radiation can override the effect of Hawking evaporation leading to the net growth and longer survival of PBHs~\cite{ASM10, upad, custodio1, custodio2}. In fact, such a phenomenon could be more prominent if the very early universe undergoes a phase of string- or brane-affected
modified expansion~\cite{ASM1, guedens1, guedens2, ASM2}, or modified geometry
for compact objects~\cite{ASM4, ASM5}. The
consequent mass gain persists during the subsequent standard radiation dominated expansion, and further impacts the rate of binary formation~\cite{ASM3}. It has been recently argued that gravitational radiation due to mass variation can substantially exceed that due to orbital motion~\cite{Holgado}.

In the present work we  focus on such PBH binaries in the early universe. Our motivation is to 
explore the 
effects of background expansion as well as accretion of radiation on the PBH binary parameters
leading to modification of the emitted gravitational wave spectrum.  We investigate the consequent alteration of the stochastic gravitational wave background, which, if detected, would lead to a direct signal of physics in the early universe, and may arguably provide a proof of existence of PBHs, as well. 
The plan of the work is as follows. In the next section we present a brief overview of binary
formation by PBHs in the early universe. In section III, we discuss the formalism for calculating 
the amplitude of gravitational waves from accreting PBH binaries. The stochastic background produced
by them is computed in Section IV where we further discuss the detectability of the resultant
spectral density by present and future gravitational wave detectors. We conclude with a summary of our analysis in section V.

\section{Binary formation by PBHs in the Early Universe} \label{PQPBH}

In the early Universe the PBHs produced at time $t$ (in seconds) after Big-bang, have masses of order of the particle horizon mass at their formation epoch, given by \cite{Carr1}
\begin{equation} \label{3.11}
m_{H}(t) \approx \frac{c^{3}t}{G} \approx 10^{15}\frac{t}{10^{-23} } \, g = 10^{38}t \, g  .
\end{equation}
Hence, PBHs may span an enormous mass range from the end of inflation ($ 10^{-32} $ s) up to the Big-bang nucleosynthesis ($\sim$ 1 s) \cite{Carr4}.
 Till the time t$ \approx \, 10^{-25}$ s, almost all  PBHs  have mass-range ($< 10^{13}$ g) such that the Hawking evaporation is dominant over the accretion of radiation for them, leading to decrease in their mass. PBHs produced after $> 10^{-25} $ s, undergo mass gain by accretion of the highly dense radiation which dominates over the Hawking evaporation. 

It has been argued earlier \cite{Carr2, Green} that if the primordial fluctuations obey a Gaussian distribution,  the probability, that a collapsing spherical region of initial mass $m$ has a density contrast in the range $\delta $ and $\delta + d\delta $, is given by \cite{Press} 
\begin{equation} \label{3.12}
P(\delta, m) d\delta = \frac{1}{\sqrt{2\pi} \sigma(m)} exp.\Big(- \frac{\delta^{2}}{2 \sigma(m)^{2}}   \Big) \Theta \Big[ \alpha \Big( \frac{m}{m_{0}} \Big)^{2/3} - \delta  \Big]  d\delta  \, ,
\end{equation}
where $\sigma(m) $ is the mass variance and  $\Theta$ denotes the Heaviside function, which represents the fact that the distribution is cut off above a maximum value of the density contrast $\delta_{\max}$. 
The cumulative density of PBHs at time $t$  is given by \cite{Carr2} 
\begin{equation} \label{3.13}
\rho_{PBH} = \int_{m_{min}}^{m_{max}} \Pi(m') \rho \, dm'    \, .
\end{equation}
where $\rho $ is the radiation density and $ \Pi(m)dm $ is a quantity related to the probability that a spherical region having initial mass between $m$ to $m+dm$ collapses to a PBH, which ultimately remains a single black hole, i.e., is not engulfed by any larger black hole \cite{Carr2}. Since in the present work we are interested in
those PBHs for which the mass gain due to accretion of radiation is dominant over mass loss due to Hawking evaporation,
we set the limits of the above integral to be $m_{min} = 10^{13}$ g and $ m_{max} = m_{H}$.
The  mass-variance $\sigma(m)$ is taken as $\sigma(m) = \epsilon ( m/m_{0} ) ^{-n}$, where $ \epsilon$ is a  constant, $ m_{0} $ is the initial sub-horizon mass and  $n = 2/3 $~\cite{Carr2}.    
Hence, $\Pi(m) $ can be obtained as
\begin{equation} \label{3.14}
\Pi(m) \sim \frac{1}{m} \epsilon  \Big[ exp \Big( -\frac{\mathscr{B}^{4}}{2 \epsilon^{2}}   \Big) \Big]   \, ,
\end{equation} 
where, $\mathscr{B}^{2} \sim w $, with $w $ being the equation of state parameter of the concerned cosmic-fluid, which is radiation in this case. 

Binary formation of PBHs in the early Universe typically proceeds due to the decoupling of a pair of PBHs
from the background cosmic expansion, with a third nearby PBH providing a tidal force to prevent
head-on collision~\cite{naka1, naka2, ASM3}.  The scale-factor at which a pair of PBHs decouple from the cosmic expansion is given by~\cite{Raidal} 
\begin{equation}   \label{6.9.1}
a_{dc} \approx a_{eq} \Big( \frac{r_{dc}}{\tilde{r}} \Big)^{3}  \, , 
\end{equation} 
where, $r_{dc}$ is the co-moving separation between the two black holes, $\tilde{r} $ is given by 
\begin{equation}   \label{6.9.1e}
\tilde{r}^{3} = \frac{3}{4\pi} \frac{M}{a_{eq}^{3} \rho_{eq}}
\end{equation}   
 and $M=m_{1}+m_{2} $ is the total mass of the two PBHs decoupling from cosmic-expansion. $a_{eq} $ and $\rho_{eq} $ are respectively the scale-factor and density of cosmic-fluid at the matter-radiation equality. 
As the concerned era is  radiation dominated, the corresponding time is given by 
\begin{equation}   \label{6.9.2}
t_{dc} = \mathscr{A}^{-2} a_{dc}^{2} \approx  \mathscr{A}^{-2} a_{eq}^{2} \Big( \frac{r_{dc}}{\tilde{r}} \Big)^{6}  \, , 
\end{equation}     
where $ \mathscr{A} $ is a constant defined by $a =  \mathscr{A} t^{1/2} $. 
Here, the co-moving length-scale $\tilde{r}$ comes from the condition of decoupling from cosmic expansion, which is roughly when the mean mass of the pair of PBHs overtakes the mass of the cosmic fluid contained in the sphere of radius equal to the separation of the pair, given by
\begin{equation}  \label{6.9.3}
\frac{M}{2} > \frac{4\pi}{3 c^{2}} \rho R^{3}    \,   , 
\end{equation}
where $R$ is the proper separation between the PBHs and $\rho$ is the density of the cosmic-fluid i.e. radiation. As  argued in the references 
~\cite{Sasaki_et_al} and ~\cite{Raidal}, the length-scale $\tilde{r}$ is such that  $r_{dc}  < \tilde{r}$ and consequently $a_{dc} <  a_{eq} $. 
The decoupling time must be greater than the time of production of both the PBHs, given by equation (\ref{3.11}), viz. $t \approx \frac{G}{c^{3}} m $. Hence, the combination of masses $m_{1}$ and $m_{2}$ in the total mass $M$ (on which $t_{dc}$ depends) should not be such that one of them is very large and the other is very small making the time of formation of the larger PBH greater than $t_{dc}$. 

In the early inspiral stage, the angular-frequency of the PBH-binaries (just after formation of the binary) is given by : $\omega = (G(m_{1}+m_{2})/R_{dc}^{3})^{1/2} $, where substituting the expression of proper separation between them : $R_{dc} \approx \tilde{r} \frac{a_{dc}^{4/3}}{a_{eq}^{1/3} }    $, we can obtain : 
\begin{equation}  \label{6.9.3e}
\omega =  \Big(  \frac{4 \pi G(m_{1}+ m_{2})   a_{eq}^{4} \rho_{eq}}{3 \mathscr{A}^{4} t_{dc}^{2} (m_{1}+ m_{2})  } \Big)^{1/2} = \Big( \frac{4 \pi G a_{eq}^{4} \rho_{eq}}{ 3 \mathscr{A}^{4} t_{dc}^{2} } \Big)^{1/2}    
\end{equation}   

The radial distance of a PBH-binary at a scale-factor 'a' is given by the usual formula for cosmological distance, 
\begin{equation} \label{6.9.4}
 D(a) = \frac{c}{H_{0}} \int_{a}^{1}   \frac{da}{a^{2}(\Omega_{DE} +  \Omega_{M} a^{-3} )^{1/2}}  \,  , 
\end{equation}
where, $\Omega_{DE}$ and $\Omega_{M}$ are the fractional densities of dark energy and non-relativistic matter at present Universe and we have neglected the fractional density of radiation at present Universe $\Omega_{R}$, as $\Omega_{R} \, <<\, \Omega_{DE}, \Omega_{M}  $. 
The distance $D$ of a PBH-binary is dependent on the masses of the PBHs constituting the binary since the time of binary formation depends on the total mass of the two PBHs and also on the initial comoving-separation between the PBHs after forming the binary.
Substituting the values of $\Omega_{DE}$ and $\Omega_{M}$ and performing the integration, one obtains the distance $D(m_{1}, m_{2}, r_{dc})$ to a specific PBH-binary, given by 
\begin{equation} \label{6.9.7}
D(a) \approx \frac{c}{H_{0}}  \Big(-\xi_{1} + \frac{\xi_{2}  \, _{2} \mathcal{F}_{1}(\frac{1}{3}, \frac{1}{2}, \frac{4}{3}, - \frac{\zeta}{a^{3}})}{a}   \Big)    .
\end{equation}
Here, $_{2} \mathcal{F}_{1}$ is the Hypergeometric function, and
 $\xi_{1} \, , \xi_{2} $ and $  \zeta $ are quantities whose numerical values depend on $ \Omega_{DE} $ and $ \Omega_{M} $.  
 
Before proceeding further, it may be pertinent to mention certain observational constraints on
the abundance of PBHs in particular mass ranges~\cite{Carr_Kohri_et_al,Barnacka_et_al,Capela_et_al}.  From the absence of noticeable microlensing,   the ERS and MACHO surveys have excluded large abundances of PBHs in the mass-range $10^{26} $ to $10^{34} $ g ~\cite{Tisserand_et_al}\cite{Alcock_et_al}\cite{Griest_et_al}. This constraint can be indirectly applied to the abundance of PBHs of masses  $< \, 10^{26} $ g in the early Universe. However, setting these mass limits are highly model-dependent~\cite{Clesse_et_al_3} and regrouping of PBHs in dense halos  can evade the microlensing constrains. The absence of some characteristic spectral distortions of the Cosmic Microwave Background's spectrum imposes constraints on the PBH abundance in the 
early Universe. Planck observations exclude PBHs of the mass-order $10^{35} - 10^{37} $ g from being 
a significant fraction of dark-matter \cite{Ali-Haimoud_et_al}. However, the distortion constraints are results of complex processes and  subject to considerable uncertainties. Moreover, it has been argued~\cite{Clesse_et_al} that the rate of binary formation and consequent merging of PBHs in the early Universe could be significantly higher such that the PBHs produced with sub-stellar masses, bypassing the CMB-distortion constrains, would have grown by several orders of mass by the time of star formation. Hence, such PBHs could evade the most stringent micro-lensing constrains, as well.


\section{Gravitational wave amplitude from accreting PBH binaries}

 The second mass-moment, or quadrupole-moment, as it is known for the Transverse-Traceless (TT) gauge, of  a binary compact object  is given by  ${\displaystyle I_{ij} \equiv  \int \rho x_{i}x_{j} = \mu xy }$, for cross-polarization,
 where  the orbital-plane of the binary is chosen to be the $xy$-plane with origin at their center-of-mass, and it is assumed for simplicity that the $z$-axis is along the line from the center of the binary to the observer.  
So, if $\mu$ is constant,  ${\displaystyle {\dot{I}_{ij}}= \frac{d}{dt} \int d^{3}x (\rho(t,x)x_{i}x_{j})= \frac{d}{dt}  (\mu x(t)y(t)) }$,
which gives 
${\displaystyle  {\ddot{I}_{ij}}= \mu (\frac{d^{2}x}{dt^{2}}y+x\frac{d^{2}y}{dt^{2}})+ 2\mu (\frac{dx}{dt}\frac{dy}{dt})   }    $.
However, if $ \mu $ varies with time, there would be two extra terms, i.e., ${\displaystyle {\ddot{I}_{ij}}= \frac{d^{2}\mu}{dt^{2}}xy + \frac{d\mu}{dt}(\frac{dx}{dt}y +x \frac{dy}{dt})+}$
${\displaystyle \mu (\frac{d^{2}x}{dt^{2}}y+x\frac{d^{2}y}{dt^{2}})+ 2\mu (\frac{dx}{dt}\frac{dy}{dt}) }$.
Hence, the gravitational wave amplitude in cross($ \times $)-polarization from a single PBH-binary of continuously changing PBH masses $m_{1} $ and $m_{2} $, in the early inspiral stage where the Keplarian-approximations are valid,  will be given by
\begin{equation}  \label{6.9}
\begin{aligned}
  h_{\times} =  \frac{2G}{D\, c^{4}}{\ddot{I}_{xy}}           
  =\frac{G^{\frac{5}{3}}}{D(m_{1}, m_{2})\, c^{4}} \Big[\frac{m_{1}m_{2}}{{(m_{1}+m_{2})}^{1/3}}    \lbrace  -4\omega^{\frac{2}{3}} {sin(2\omega t)}  \rbrace  
  \\
+  \Big\lbrace    \frac{d^{2}}{dt^{2}} \frac{m_{1}m_{2}}{{(m_{1}+m_{2})}^{1/3}} \Big\rbrace {\omega^{-\frac{4}{3}}} sin(2\omega t) 
\\  
 +   2 \Big \lbrace\frac{d}{dt}\frac{m_{1}m_{2}}{{(m_{1}+m_{2})}^{1/3}}  \Big \rbrace \lbrace  2\omega^{-\frac{1}{3}} 
  cos (2\omega t)   \rbrace  \Big] 
\end{aligned}  .
\end{equation}  
(the terms generated due to time-variation of angular frequency are negligible here.)
In the RHS of the equation (\ref{6.9}) the first term is the usual one for binaries of constant mass black holes, while the rest two are present if the masses of the black holes in the binary  change with time. 

The time-rate of change of mass of any non-rotating PBH in early Universe, due to spherical
 accretion of the surrounding radiation is approximately given by,  
\begin{equation} \label{6.17}
\dot{m} = 4\pi \mathcal{A} \Big(\frac{Gm}{c^{2}}\Big)^{2} (1+w)\rho   \, 
\end{equation}
where the constant $ \mathcal{A}$ is proportional to the energy flux going into the black hole. To simplify matters, we take $\mathcal{A} = 4$, considering the Schwarzschild
cross-section for absorbtion of radiation  \cite{ASM1, ASM2}. This time rate of change of mass of a PBH is also valid when it is in the early inspiral stage of a binary.

In order to have an idea of the rate of mass gain during the radiation dominated era, we plot
$\dot{m}$ versus time for a range of PBH formation masses $m_{H} $ in Fig. \ref{mdot_vs_time}. One sees that $\dot{m}$ can indeed take large values during the early radiation dominated era, but falls rapidly with time. This is due to the fall in background radiation density.\\ 
Next, using the Friedmann's equations of FLRW-cosmology $H^{2} = \frac{8 \pi G}{3 c^{2}} \rho$ and the conservation equations of energy-momentum tensor of the cosmic fluid, {\it viz}. $\dot{\rho} = -3 H (1+w) \rho $ and substituting in Eq.(\ref{6.17}), after taking its derivative, we get 
\begin{equation}  \label{6.22}
\begin{aligned}
\ddot{m} = \frac{4\pi \mathcal{A}G^{2}}{c^{4}}(1+w)   \Big[ -3\Big(\frac{8 \pi G}{3 c^{2}}\Big)^{1/2} m^{2} (1+w) \rho^{3/2} +                          \\                       
  2  \Big( 4\pi \mathcal{A} \Big(\frac{G}{c^{2}}\Big)^{2} (1+w) m^{3} \Big) \rho^{2} \Big]  \, .
\end{aligned}  
\end{equation}
A plot of $-\ddot{m}$ versus time in Fig. \ref{mddot_vs_time} reveals that the nature of variation of $-\ddot{m} $ with time is quite similar to that of the variation of $\dot{m} $ with time. It starts from huge values during the early radiation dominated era, while falls rapidly with time. The reason, for $\ddot{m} $ having negative values, is clearly the fall of $\frac{dm}{dt} $ with time due to decreasing background radiation density. 

\begin{widetext} 
\includegraphics[width=16.3cm]{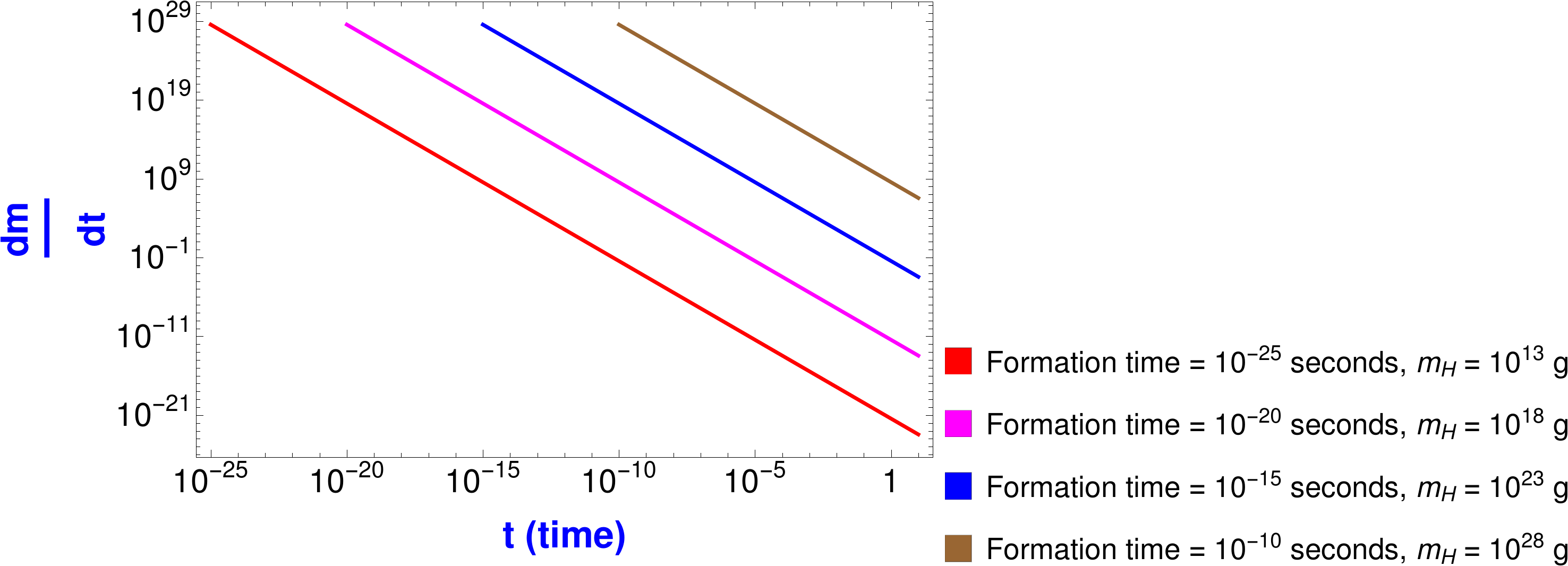}
  \captionof{figure}{ Plot of $\frac{dm}{dt}$ vs time t, where $ \frac{dm}{dt} $ is in units of $g/s$ and time t is in seconds after Big-bang. We have plotted a family of four curves for four different initial masses viz. $ m_{H}$ with the values $10^{28}, \, 10^{23}, \, 10^{18} $ and $ 10^{13} $ g.  }\label{mdot_vs_time} 
 \includegraphics[width=17.7cm]{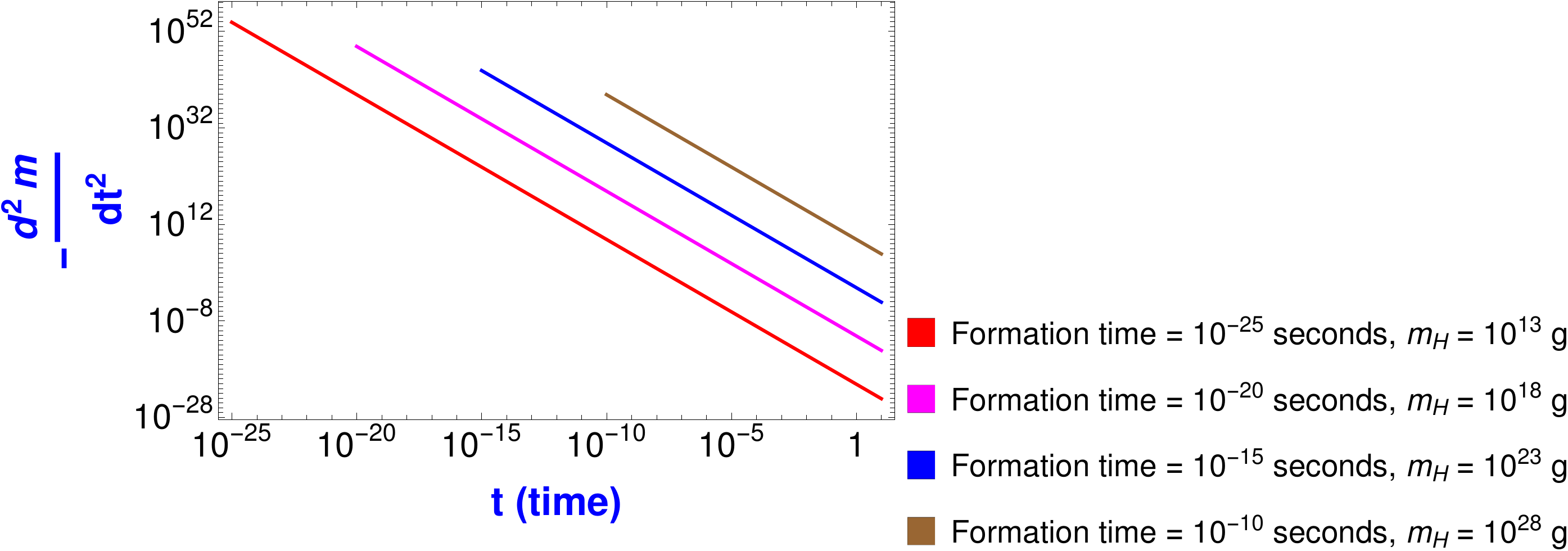}
 \captionof{figure}{Plot of $-\frac{d^{2}m}{dt^{2}}$ vs time t, where $ \frac{d^{2}m}{dt^{2}} $ is in units of $g/s^{2} $ and time t is in seconds after Big-bang. We have plotted a family of four curves for four different initial masses viz. $ m_{H}$ with the values $10^{28}, \, 10^{23}, \, 10^{18} $ and $ 10^{13} $ g. } \label{mddot_vs_time}

\includegraphics[width=17.3cm]{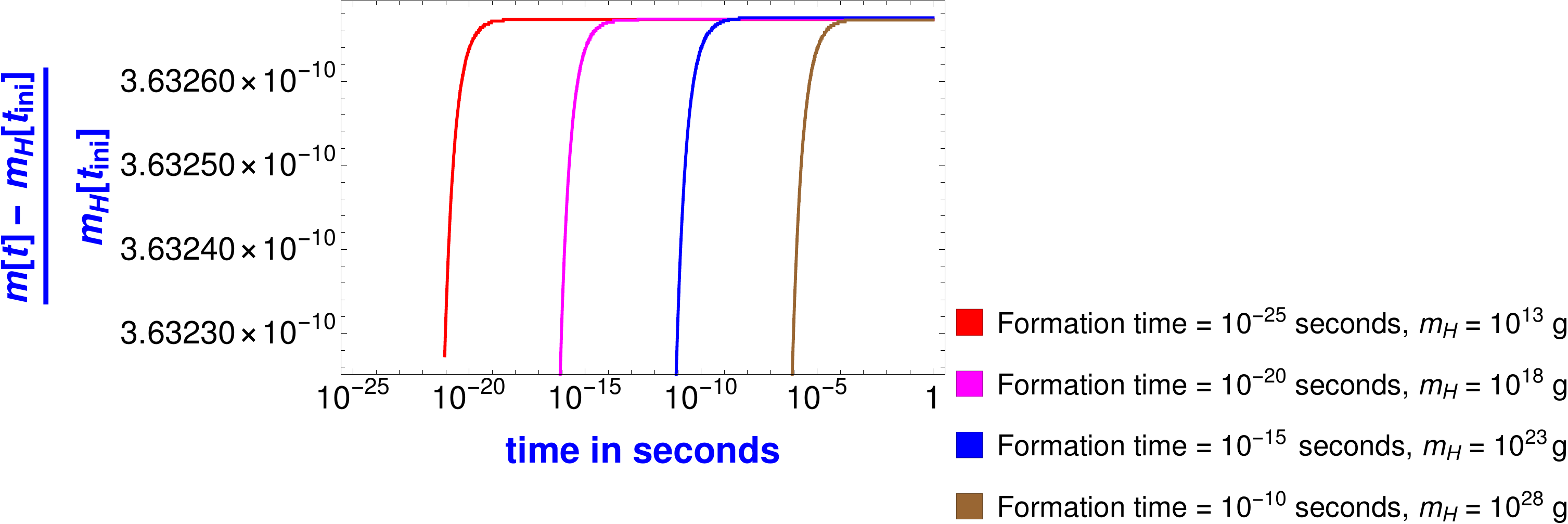}
\captionof{figure}{Plot of $\frac{m(t)- m_{H}}{m_{H}} $, which is the ratio of the growth in mass of a PBH, to its initial mass, w.r.t. time. The range of time shown in the figure is from $ 10^{-25} $ to $ 1 $ s after Big-bang. We have plotted a family of four curves for four different initial masses viz. $ m_{H}$ with the values $10^{28}, \, 10^{23}, \, 10^{18} $ and $ 10^{13} $ g. } \label{PBHmassgrowth}
\end{widetext}
Next, in Fig. \ref{PBHmassgrowth} we give a plot of the time-variation of the ratio of change in mass of a PBH taken to its initial mass, with which it was born i.e. the horizon mass $m_{H} $. It is evident from this figure that the growth of the PBHs, for the specified range of initial masses, are negligigle in comparison with their initial masses. The amount of growth of the PBHs' masses are less than of the order of $ 10^{-12} $ times of their initial masses, in the range of time of our interest. Hence, it is in clear agreement with the argument of B. J. Carr and S. W. Hawking in their work \cite{HawkingCarr} that PBHs can not grow much significantly in the radiation dominated era. Various other works also suggest the same \cite{Bicknell1, Bicknell2}. It can also be noticed from this figure \ref{PBHmassgrowth}, that initially the masses grow faster for a little time, after which they tend to become constant. The reason behind this can be interpreted as the rapid fall in the background radiation density in this early radiation-dominated era, due to which the rate of growth of PBH masses also fall rapidly with the evolution of Universe. So, it is very interesting to note the fact that although the growth of masses of the PBHs are negligible when we compare that with their initial masses, but yet the rate of growth is sufficient to have a significant impact on the gravitational wave emitted from their binaries, which we shall show in our work. 
\\
Now, it is to be noted that both the $\dot{m} $ and $\ddot{m} $ have been expressed in terms of $m $ and $\rho $,
enabling one to write the single and double time derivatives of the chirp-mass function, in the correction terms in cross-polarization of gravitational wave amplitude in Eq.(\ref{6.9}), respectively as :
\begin{equation}   \label{6.23}
\begin{aligned}
\frac{d}{dt}\frac{m_{1}m_{2}}{{(m_{1}+m_{2})}^{1/3}}  =                  \\
 \Big \lbrace  \frac{4\pi \mathcal{A} G^{2} }{c^{4}}(1+w)\rho   \Big \rbrace \Big[  m_{1} m_{2}(m_{1}+m_{2})^{2/3}   -  \frac{m_{1}m_{2}(m_{1}^{2}+m_{2}^{2})}{3(m_{1}+m_{2})^{4/3}}  \Big]
\end{aligned}
\end{equation}   
and
\begin{equation}   \label{6.24}
\begin{aligned}
  \frac{d^{2}}{dt^{2}}\frac{m_{1}m_{2}}{{(m_{1}+m_{2})}^{1/3}}  =  \frac{ 4 \pi \mathcal{A} G^{2} }{c^{4}}(1+w) \rho \frac{m_{1}^{2}m_{2}^{2}}{(m_{1}+m_{2})^{1/3}} +                   \\
  \frac{4\pi \mathcal{A}G^{2}}{c^{4}}(1+w)^{2} \Big( \frac{(\mathscr{C}_{1} m_{1}^{2} \rho^{3/2} + \mathscr{C}_{2} m_{1}^{3} \rho^{2} ) m_{2}} {(m_{1}+m_{2})^{1/3}} +        \\
  \frac{ (\mathscr{C}_{1} m_{2}^{2} \rho^{3/2} + \mathscr{C}_{2} m_{2}^{3} \rho^{2} ) m_{1} }              {(m_{1}+m_{2})^{1/3}}   \Big)                                              \\
  - \Big(  \frac{4\pi \mathcal{A}G^{2}}{c^{4}}(1+w)\rho  \Big)^{2} \frac{m_{1}m_{2}(m_{1}+ m_{2})(m_{1}^{2}+ m_{2}^{2})}{(m_{1}+ m_{2})^{4/3}}                                           \\
- \frac{4\pi \mathcal{A}G^{2}}{c^{4}}(1+w)^{2} \frac{m_{1}m_{2}}{3(m_{1}+m_{2})^{4/3}}   ( \mathscr{C}_{1} \rho^{3/2} ( m_{1}^{2} + m_{2}^{2} ) +\\    \mathscr{C}_{2}  \rho^{2} (m_{1}^{3} + m_{2}^{3}) )      
+  \frac{4\pi \mathcal{A}G^{2}}{c^{4}}(1+w)\frac{4m_{1}m_{2} (m_{1}^{2} + m_{2}^{2} )^{2}}{9(m_{1}+ m_{2})^{7/3} }  \,   , 
\end{aligned}   
\end{equation}
where the quantities $\mathscr{C}_{1} $ and $\mathscr{C}_{2} $ are, respectively, $-3(8 \pi G/ 3 c^{2})^{1/2} $ and $8 \pi \mathcal{A} (G/c^{2})^{2} $. 
We can now calculate the numerical values of the peak magnitudes (without the sinusoidal variations) of the first and second corrections terms in gravitational wave amplitude given by $\frac{G^{5/3}}{D\, c^{4}} \Big\lbrace    \frac{d^{2}}{dt^{2}} \frac{m_{1}m_{2}}{{(m_{1}+m_{2})}^{1/3}} \Big\rbrace {\omega^{-\frac{4}{3}}}  $, and  
  $ \frac{G^{5/3}}{D\, c^{4}} 2 \Big \lbrace\frac{d}{dt}\frac{m_{1}m_{2}}{{(m_{1}+m_{2})}^{1/3}}  \Big \rbrace \lbrace  2\omega^{-\frac{1}{3}}    \rbrace  $ respectively, for any typical PBH binary and compare their values with that of the main term $ \frac{G^{5/3}}{D\, c^{4}} \frac{m_{1}m_{2}}{{(m_{1}+m_{2})}^{1/3}} \lbrace  -4\omega^{\frac{2}{3}} \rbrace  $. \\
  We plot these terms in Fig. \ref{CompInd} as functions of the black hole masses and the background radiation density, choosing $m_{2} = 2 m_{1}$ and  separation between the PBHs is given by $100$ times the sum of their Schwarzschild-radii (the angular frequency is to be directly obtained from Kepler's law as we are considering the early inspiral
stage).  We  use the expression for cosmological distance in terms of the scale factor given by equation \ref{6.9.7}, considering the scale factor at which the PBHs constituting the binary were born (as masses of both the PBHs are of same order, their time of birth is also of approximately same order). It is evident from the plot that for certain cases the corrections are not only significant but also dominant. The constancy of the main term w.r.t. the background radiation density $\rho $ can be clearly depicted in the plot below, as it is independent of $ \rho $. With the increasing density of radiation $ \rho $, both the correction terms increase. Therefore, the instantaneous rate of change of masses (both the single and double time-derivatives of the masses) of the PBHs in binaries have a significant effect on the gravitational wave amplitude generated by them and hence, on the overall stochastic gravitational wave background.
\begin{widetext}
\begin{center}
\begin{figure}
\includegraphics[width=17.0cm]{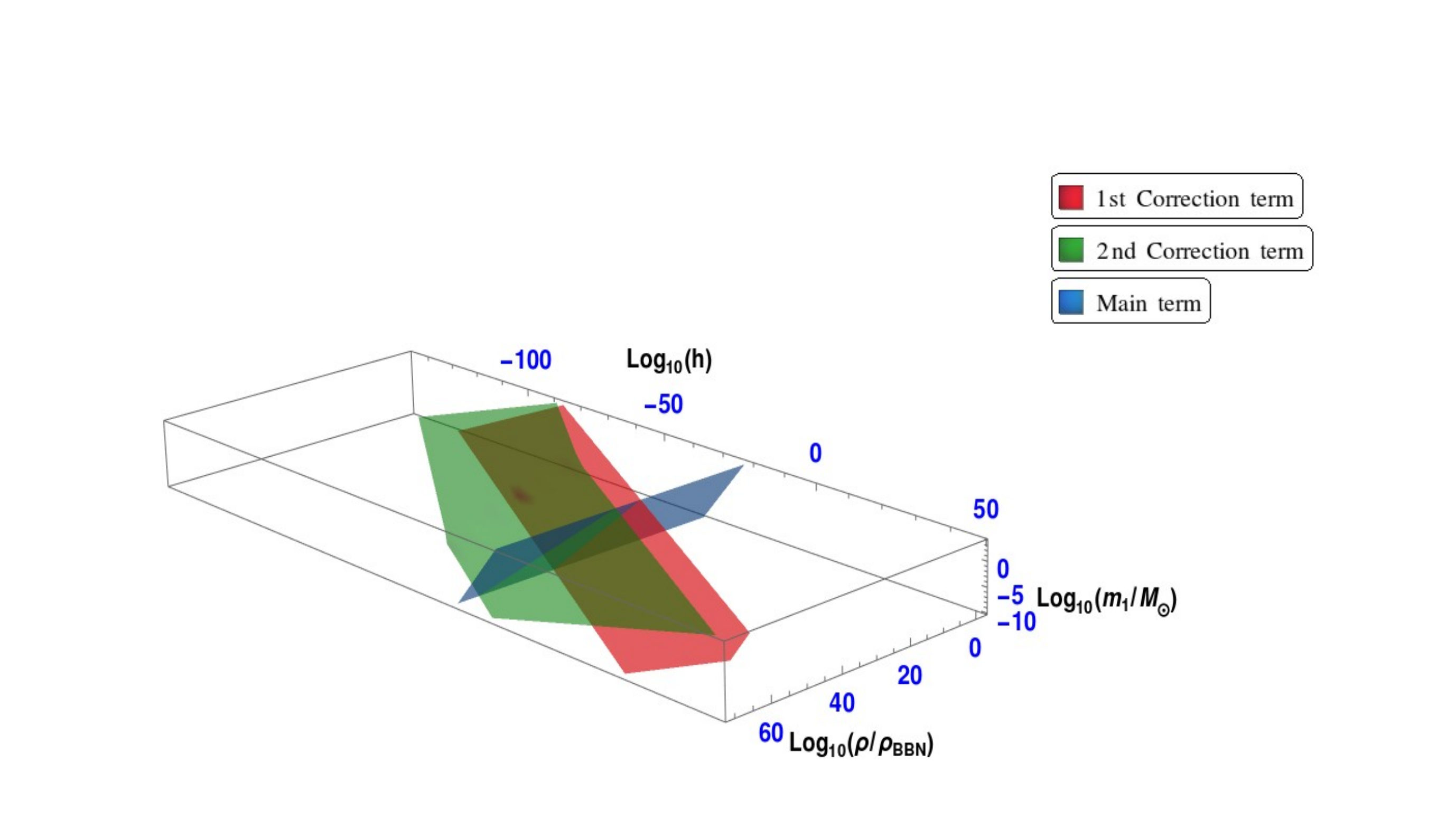}
 \caption{Variation of the three terms in the gravitational wave amplitude given by Eq.(\ref{6.9}) produced by a PBH binary in the early inspiral stage, w.r.t. the  background radiation density and mass, where $ \rho_{BBN} \approx 10 g/cm^{3}$ is the radiation density during big-bang nucleosynthesis, and
$M_{\odot} $ is the Solar-mass.  
    }\label{CompInd} 
\end{figure}
\end{center}
\end{widetext}
\section{Stochastic gravitational wave background and its detectability}\label{ParaSGWB}
For calculating the stochastic background, we employ the standard formalism ~\cite{Maggiore}  assuming
that the background is stationary, Gaussian, isotropic and unpolarized. Under these assumptions,
the  spectral density of the stochastic background $S_{h}(f)$ is defined as :    
\begin{equation} \label{6.1a}
S_{h}(f) = \frac{1}{4} \frac{d}{df} \langle h_{ij}(t)h^{ij}(t) \rangle   \,  ,  
\end{equation}
where $h_{ij}(t) \equiv h_{ij}(t, \vec{r}=0) $. The brackets $ \langle$ $ \rangle $ over the scalar product $ h_{ij}(t)h^{ij}(t) $ denote in this case the average taken over certain interval of 
time~\cite{Maggiore}. 
The advantage of describing the theory in terms of the spectral density $S_{h}(f) $ is that it is directly comparable with the noise in a detector, denoted by $S_{n}(f) $. 
 The response to any stochastic gravitational wave background by the detector is given by
\begin{center}
$ {\displaystyle  h(t) = \frac{F}{4} \langle  h_{11}h^{11} +  h_{12}h^{12} +  h_{21}h^{21} +  h_{22}h^{22}  \rangle^{1/2}  }$\\
$ {\displaystyle = \frac{F}{4} \langle  h_{11}h^{11} + 2 h_{12}h^{12} +   h_{22}h^{22}  \rangle^{1/2}   } \,   , $   
\end{center} 
   given in terms of the spectral density by 
 \begin{equation} \label{6.e2}
h(t) = \langle  h^{2}(t) \rangle^{1/2} =  \Big( F \int_{f} df S_{h}(f) \Big)^{1/2} =\Big(  \frac{F}{4} \langle h_{ij}(t)h^{ij}(t)  \rangle \Big)^{1/2}   \, ,
 \end{equation}
 where $F$ is the angular efficiency factor, which  for interferometric detectors is $F= 2/5 $, and for cylindrical bar detectors $F=8/15 $ ~\cite{Maggiore}.  
 
Substituting the expressions of plus and cross polarized components of gravitational wave amplitude in the expression of $h(t)$ one gets,
\begin{equation} \label{6.e3}
\begin{aligned}
     h(t) =   \frac{F}{4}  \frac{2 \, G^{\frac{5}{3}}}{D(m_{1}, m_{2}) \, c^{4}}  
   \Big[ 2\Big(2 \frac{m_{1}m_{2}}{{(m_{1}+m_{2})}^{1/3}}   \omega^{\frac{2}{3}}   \Big)^{2} + 
 \\                        
   2 \Big( 2 \Big \lbrace\frac{d}{dt}\frac{m_{1}m_{2}}{{(m_{1}+m_{2})}^{1/3}}  \Big \rbrace   \omega^{-\frac{1}{3}}    \Big)^{2}   +               
  \Big(   \Big\lbrace    \frac{d^{2}}{dt^{2}} \frac{m_{1}m_{2}}{{(m_{1}+m_{2})}^{1/3}} \Big\rbrace {\omega^{-\frac{4}{3}}}     \Big)^{2}              +           \\
    4   \Big\lbrace    \frac{d^{2}}{dt^{2}} \frac{m_{1}m_{2}}{{(m_{1}+m_{2})}^{1/3}} \Big\rbrace {\omega^{-\frac{4}{3}}}  \Big( -2\frac{m_{1}m_{2}}{{(m_{1}+m_{2})}^{1/3}}   \omega^{\frac{2}{3}}  \Big)        \Big]^{1/2} \,  .  
\end{aligned}  
\end{equation} 
An additional cross-correction term appears due to the non-vanishing of the product between the main term and the second correction term, i.e. the term containing the double-time derivative of chirp-mass. Unlike the case of a single binary, $h_{ij} $ for stochastic background of gravitational waves stands for the overall gravitational wave amplitude of the stochastic background, integrated over all possible frequencies and all directions, given by 
\begin{equation} \label{6.6}
\begin{aligned}
h_{ij}(t,\vec{r}) = \underset{\mathcal{P}= +,\times}{\Sigma} \int_{f} df  \int d^{2}\hat{n}   \\
 h_{\mathcal{P}}(f,\hat{n},t) e^{\mathcal{P}}_{ij}(\hat{n}) exp[-2\pi i f(t- \hat{n}.\vec{r}/c)]   \,  ,  
\end{aligned}
\end{equation}
where $h_{\mathcal{P}} $ is the gravitational wave amplitude produced from each PBH binary. 
The constituent gravitational waves from all the PBH-binaries come from all directions or the overall solid angle, as they are statistically distributed in the Universe and with statistically distributed parameters. Hence, for
 the stochastic background,  one has to integrate over all the solid angles and over their masses.

Denoting the chirp masses as 
$ \mathcal{M}= \frac{(m_{1}m_{2})^{3/5}}{(m_{1}+m_{2})^{1/5}}   $, 
 the density of gravitational wave amplitude generated from PBH binaries in the differential chirp-mass range $\mathcal{M} $ to $\mathcal{M}+ d \mathcal{M}$, for cross-polarization is given by 
\begin{equation} \label{6.12}
dh_{ij}(t,\vec{r}) =   \int_{f} df  \int d^{2}\hat{n} \, \mathcal{N}(\mathcal{M})d\mathcal{M} \, (h_{\times}  e^{\times}_{ij}(\hat{n}) )    \,   , 
\end{equation}   
where $ \mathcal{N}(\mathcal{M})d\mathcal{M} $ is the number-density of PBH-binaries in the differential chirp-mass range $\mathcal{M} $ to $\mathcal{M}+ d \mathcal{M}$ at the concerned time. The total gravitational wave amplitude density generated for cross-polarization from all the PBH binaries in the chirp-mass range from $\mathcal{M}_{min} $ to $\mathcal{M}_{max} $ is given by
\begin{equation}   \label{6.13}
\int dh_{ij}(t,\vec{r}) =   \int_{f} df  \int d^{2}\hat{n}  \int_{\mathcal{M}_{min}}^{\mathcal{M}_{max}}  \mathcal{N}(\mathcal{M})d\mathcal{M} (h_{\times}  e^{\times}_{ij}(\hat{n}))    \,  . 
\end{equation}
The formation of binaries is taken to proceed under the three-body configuration~\cite{Raidal}.  The differential co-moving number-density of PBH-binaries resulting from  three-body configurations may be 
written as
\begin{equation} \label{7.1}
d\mathcal{N}(r_{1}, r_{2}) = \frac{1}{2} (n(m_{1})dm_{1}) (e^{-N(r_{2})} dN(r_{1}, m_{2}) dN(r_{2}, m_{3}) )  \, .
\end{equation}
The part $ (e^{-N(r_{2})} dN(r_{1}, m_{2}) dN(r_{2}, m_{3}) )  $  stands for the probability that those PBHs belong to the specified three-body configuration. The quantity $dN(r,m)$ is given by
\begin{equation} \label{7.2}
dN(r,m) = 4\pi r^{2} n(m) (1+\xi(r)) dr dm    \,  .
\end{equation}
Here, $\xi(r)$ is the PBH two-point function ~\cite{Raidal}. In the simplest case, the two-point function can be taken as a constant $(1+\xi(r))= \delta_{dc} $. The factor $1/2 $ in the RHS of equation \ref{7.1} signifies the fact that the number of PBH-binaries would be just the half of the number of PBHs forming those binaries and $N(r_{2}) = \int dN(r_{2},m)$ is the expected number of PBHs surrounding one PBH in the sphere of co-moving radius $r_{2}$.
The quantity $N(r_{2}) $ is given by 
\begin{center}
 ${\displaystyle N(r_{2}) = \int_{0}^{r_{2}} \int_{m_{min}}^{m} \delta_{dc} (4\pi r^{2} dr) \Big(  \rho \frac{\Pi(m)}{m}   dm \Big)  }   \,  . $    
 \end{center}
Substituting the expression of the distribution function given by Eq.(\ref{3.14}), and performing the integrals over $r$ and $m$ in $ N(r_{2}) $, one gets
 \begin{center}
${\displaystyle  N(r_{2}) =  \delta_{dc} \rho \Big(\frac{4}{3} \pi r_{2}^{3} \Big) \epsilon \, exp.\big( \frac{-w^{2}}{2 \epsilon^{2}}  \big)\Big( -\frac{1}{m} + \frac{1}{m_{min}}  \Big)    }  \,  .$
\end{center}
For brevity of notation we define $\mathscr{N}(m) $ as $N(r_{2}) = \frac{4}{3}\pi r_{2}^{3} \mathscr{N}(m) $.\\

The total gravitational wave amplitude density generated (for cross-polarization)
from all the PBH-binaries is given by 
\begin{equation} \label{7.3}
\begin{aligned}
h_{ij}(t,\vec{r})  =  \int dh_{ij}(t,\vec{r}) =   \int_{f} df  \int d^{2}\hat{n}  \\  
\int_{m_{1,min}}^{m_{1, max}}   \int_{m_{2, min}}^{m_{2, max}}  \int_{m_{3, min}}^{m_{3, max}} \int_{r_{1}=0}^{\tilde{r}} \int_{r_{2}= r_{1}}^{\infty} d\mathcal{N}(r_{1}, r_{2})  (h_{\times}  e^{\times}_{ij}(\hat{n}))  \,  , 
\end{aligned}
\end{equation}  
where $\tilde{r}$ is defined earlier in Eq.(\ref{6.9.1e}).   
Here, the integrations over  $r_{1}$ and $r_{2}$ are  respectively from $0$ to $\tilde{r}$, and $r_{1}$ to $\infty $, because the second PBH should be within a radial distance $0$ to $\tilde{r}$ from the first PBH, while the 
third PBH has to be anywhere outside $r_{1}$ ($r_{2}> r_{1}$).
The contribution of the main term (i.e. the term without derivatives of masses) to the gravitational wave amplitude density is :
\begin{center}
${\displaystyle h_{ij \, main}(t,\vec{r})  =   \frac{1}{2} \rho^{3} \delta_{dc}^{2} \underset{f}{\int}  df  \underset{\hat{n}}{\int} d^{2}\hat{n} \int_{m_{1,min}}^{m_{1, max}}   \int_{m_{2, min}}^{m_{2, max}}  \int_{m_{3, min}}^{m_{3, max}}     }$\\
${\displaystyle \Big( \frac{\Pi(m_{1})}{m_{1}} dm_{1}  \frac{\Pi(m_{2})}{m_{2}} dm_{2} \frac{\Pi(m_{3})}{m_{3}} dm_{3}   \Big)   }$
${\displaystyle  \Big(  4 \pi \int_{r_{2}} r_{2}^{2} e^{-N(r_{2})} dr_{2}  \Big)  }$   
${\displaystyle  4\pi \int_{r_{1}} r_{1}^{2} dr_{1}  } $  
${\displaystyle  \frac{G^{5/3}}{D(m_{1}, m_{2}, r_{1})\, c^{4}} \frac{m_{1} m_{2}}{(m_{1}+ m_{2})^{1/3}} (-4 \omega^{2/3} sin(2\omega t))  }  \,  .$
\end{center}  
The contributions of the correction terms follow similarly. Note that the hypergeometric function contained in the expression of the distance $D(m_{1}, m_{2}, r_{1})$ given by Eq.(\ref{6.9.7})
can be written as 
\begin{equation}
_{2} \mathcal{F}_{1}(a,b,c,Z)  \approx \frac{\Gamma(c) \Gamma(b-a)}{\Gamma(b) \Gamma(c-a)} (-Z)^{-a} + \frac{\Gamma(c) \Gamma(a-b)}{\Gamma(a) \Gamma(c-b)} (-Z)^{-b}   \,   ,
\end{equation}  
for $|Z| >> 1$. This allows us to carry out the radial integrations analytically.
Employing the approximation of the Hypergeometric function as described above, we find that the expession of the distance can be approximately written as : 
\begin{equation}
\begin{aligned}
D(m_{1}, m_{2}, r_{1}) \approx  \mathscr{D}(1+ \alpha \, a_{dc}^{1/2})^{-1}    \,  ,     
\end{aligned} 
\end{equation}
where numerical values of $\mathscr{D} $ and $ \alpha $ are estimated to be of order $\sim $ 1.   

 The contribution from the main term in $\langle h(t)^{2} \rangle $ 
is given by
\begin{equation}  \label{10.1}
\begin{aligned}  
  \langle h(t)^{2}  \rangle_{main}  =  \frac{1}{2} \rho^{3} \delta_{dc}^{2} \underset{f}{\int}  df  \underset{\hat{n}}{\int} d^{2}\hat{n} \int_{m_{1,min}}^{m_{1, max}}   \int_{m_{2, min}}^{m_{2, max}}  \int_{m_{3, min}}^{m_{3, max}}  
  \\   
 \Big( \frac{\Pi(m_{1})}{m_{1}} dm_{1}  \frac{\Pi(m_{2})}{m_{2}} dm_{2} \frac{\Pi(m_{3})}{m_{3}} dm_{3}   \Big)   
 \Big(  4 \pi \int_{r_{2}} r_{2}^{2} e^{-N(r_{2})} dr_{2}  \Big)   \\ 
  4\pi \int_{r_{1}} r_{1}^{2} dr_{1}  
  \Big( 2\frac{F}{4}  \frac{2 \, G^{\frac{5}{3}}}{D(m_{1}, m_{2}, r_{1}) \, c^{4}} \Big)^{2}  
   \Big(2 \frac{m_{1}m_{2}}{{(m_{1}+m_{2})}^{1/3}}   \omega^{\frac{2}{3}}   \Big)^{2}       \,   ,
\end{aligned}
\end{equation}  
and similarly, for the three correction terms.

The contribution of the main term to the spectral density, after carrying  out the integrations over $ r_{2} $ and $r_{1}$, (neglecting terms containing $a_{eq} $, as the order of $a_{eq} \, << 1$) is given by,
\begin{widetext}
\begin{equation}            \label{10.4}
\begin{aligned}
S_{h}(f)_{main}  =                              \\
 \frac{1}{2} \rho^{3} \delta_{dc}^{2}  (4\pi)\Big(\frac{c}{H_{0}} \Big)^{-2}\int_{m_{1,min}}^{m_{1, max}}   \int_{m_{2, min}}^{m_{2, max}}  \int_{m_{3, min}}^{m_{3, max}}
 \Big( \frac{\Pi(m_{1})}{m_{1}} dm_{1}  \frac{\Pi(m_{2})}{m_{2}} dm_{2} \frac{\Pi(m_{3})}{m_{3}} dm_{3}   \Big) 
   \Big(2  \, \frac{G^{\frac{5}{3}}}{ c^{4}} \frac{m_{1}m_{2}}{{(m_{1}+m_{2})}^{1/3}}   \omega^{\frac{2}{3}}   \Big)^{2}  
 \\
 \frac{4\pi}{\mathscr{D}^{2} \mathscr{N}(m) } \Big[  \frac{ exp( -\frac{4}{3}\pi \, \tilde{r}^{3} \mathscr{N}(m) )}{\mathscr{N}(m)}   
\left\lbrace   -\frac{2 \alpha}{4 \pi} \, a_{eq}^{1/2} - \frac{\alpha^{2}}{4 \pi}  \right\rbrace  
 + \frac{2\sqrt{3}\alpha}{16 \pi \mathscr{N}(m)^{3/2}} \frac{a_{eq}^{1/2}}{\tilde{r}^{3/2}} (Erf[2 \sqrt{\frac{\pi}{3}} \tilde{r}^{3/2} \sqrt{\mathscr{N}(m)}]- Erf[0])]  
\frac{1}{4\pi\mathscr{N}(m)}
\Big]    \,   ,
\end{aligned}
\end{equation} 
where the $Erf[]$ denotes the error function.    
Similarly, the three corrections to the spectral density are obtained from the correction terms in $  h(t) $, {\it viz}., for ${\displaystyle 2 \Big( 2 \Big \lbrace\frac{d}{dt}\frac{m_{1}m_{2}}{{(m_{1}+m_{2})}^{1/3}}  \Big \rbrace   \omega^{-\frac{1}{3}}    \Big)^{2}   }$, 
${\displaystyle                 \Big(   \Big\lbrace    \frac{d^{2}}{dt^{2}} \frac{m_{1}m_{2}}{{(m_{1}+m_{2})}^{1/3}} \Big\rbrace {\omega^{-\frac{4}{3}}}     \Big)^{2}    }$  and 
${\displaystyle    4   \Big\lbrace    \frac{d^{2}}{dt^{2}} \frac{m_{1}m_{2}}{{(m_{1}+m_{2})}^{1/3}} \Big\rbrace {\omega^{-\frac{4}{3}}}  \Big( -2\frac{m_{1}m_{2}}{{(m_{1}+m_{2})}^{1/3}}   \omega^{\frac{2}{3}}  \Big)   }$, the contributions to $ S_{h}(f) $ are respectively :

\begin{equation}            \label{10.5}
\begin{aligned}
S_{h}(f)_{1st} =                            \\
  \frac{1}{2} \rho^{3} \delta_{dc}^{2}  (4\pi)\Big(\frac{c}{H_{0}} \Big)^{-2}\int_{m_{1,min}}^{m_{1, max}}   \int_{m_{2, min}}^{m_{2, max}}  \int_{m_{3, min}}^{m_{3, max}}   
\Big( \frac{\Pi(m_{1})}{m_{1}} dm_{1}  \frac{\Pi(m_{2})}{m_{2}} dm_{2} \frac{\Pi(m_{3})}{m_{3}} dm_{3}   \Big)
   \Big(2 \, \frac{G^{\frac{5}{3}}}{ c^{4}}  \left\lbrace  \frac{d}{dt} \frac{m_{1}m_{2}}{{(m_{1}+m_{2})}^{1/3}}  \right\rbrace  \omega^{-\frac{1}{3}}  \Big)^{2}   
 \\
\frac{4\pi}{\mathscr{D}^{2} \mathscr{N}(m) } \Big[  \frac{ exp( -\frac{4}{3}\pi \, \tilde{r}^{3} \mathscr{N}(m) )}{\mathscr{N}(m)}   
\left\lbrace   -\frac{2 \alpha}{4 \pi} \, a_{eq}^{1/2} - \frac{\alpha^{2}}{4 \pi}  \right\rbrace   
 + \frac{2\sqrt{3}\alpha}{16 \pi \mathscr{N}(m)^{3/2}} \frac{a_{eq}^{1/2}}{\tilde{r}^{3/2}} (Erf[2 \sqrt{\frac{\pi}{3}} \tilde{r}^{3/2} \sqrt{\mathscr{N}(m)}]- Erf[0])]  
 + \frac{1}{4\pi\mathscr{N}(m)}
\Big]    \,   ,
\end{aligned}
\end{equation}    
\begin{equation}            \label{10.6}
\begin{aligned}
S_{h}(f)_{2nd} =                          \\
  \frac{1}{2} \rho^{3} \delta_{dc}^{2}  (4\pi)\Big(\frac{c}{H_{0}} \Big)^{-2}\int_{m_{1,min}}^{m_{1, max}}   \int_{m_{2, min}}^{m_{2, max}}  \int_{m_{3, min}}^{m_{3, max}}   
\Big( \frac{\Pi(m_{1})}{m_{1}} dm_{1}  \frac{\Pi(m_{2})}{m_{2}} dm_{2} \frac{\Pi(m_{3})}{m_{3}} dm_{3}   \Big)
  \frac{1}{2} \Big(  \, \frac{G^{\frac{5}{3}}}{ c^{4}}    \Big\lbrace    \frac{d^{2}}{dt^{2}} \frac{m_{1}m_{2}}{{(m_{1}+m_{2})}^{1/3}} \Big\rbrace {\omega^{-\frac{4}{3}}}  \Big)^{2}   
 \\
\frac{4\pi}{\mathscr{D}^{2} \mathscr{N}(m) } \Big[  \frac{ exp( -\frac{4}{3}\pi \, \tilde{r}^{3} \mathscr{N}(m) )}{\mathscr{N}(m)}   
\left\lbrace   -\frac{2 \alpha}{4 \pi} \, a_{eq}^{1/2} - \frac{\alpha^{2}}{4 \pi}  \right\rbrace   
 + \frac{2\sqrt{3}\alpha}{16 \pi \mathscr{N}(m)^{3/2}} \frac{a_{eq}^{1/2}}{\tilde{r}^{3/2}} (Erf[2 \sqrt{\frac{\pi}{3}} \tilde{r}^{3/2} \sqrt{\mathscr{N}(m)}]- Erf[0])]  
 + \frac{1}{4\pi\mathscr{N}(m)}
\Big]      \, , 
\end{aligned}
\end{equation}    
\begin{equation}            \label{10.7}
\begin{aligned}
S_{h}(f)_{cross} =                              
  \frac{1}{2} \rho^{3} \delta_{dc}^{2}  (4\pi)\Big(\frac{c}{H_{0}} \Big)^{-2}\int_{m_{1,min}}^{m_{1, max}}   \int_{m_{2, min}}^{m_{2, max}}  \int_{m_{3, min}}^{m_{3, max}}   
\Big( \frac{\Pi(m_{1})}{m_{1}} dm_{1}  \frac{\Pi(m_{2})}{m_{2}} dm_{2} \frac{\Pi(m_{3})}{m_{3}} dm_{3}   \Big)
\\
  4\Big(  \, \frac{G^{\frac{5}{3}}}{ c^{4}}  \Big)^{2}   \Big\lbrace    \frac{d^{2}}{dt^{2}} \frac{m_{1}m_{2}}{{(m_{1}+m_{2})}^{1/3}} \Big\rbrace     
   \Big( -\frac{m_{1}m_{2}}{{(m_{1}+m_{2})}^{1/3}}  \Big)  \omega^{-\frac{2}{3}}   
   \\
\frac{4\pi}{\mathscr{D}^{2} \mathscr{N}(m) } \Big[  \frac{ exp( -\frac{4}{3}\pi \, \tilde{r}^{3} \mathscr{N}(m) )}{\mathscr{N}(m)}   
\left\lbrace   -\frac{2 \alpha}{4 \pi} \, a_{eq}^{1/2} - \frac{\alpha^{2}}{4 \pi}  \right\rbrace    
 + \frac{2\sqrt{3}\alpha}{16 \pi \mathscr{N}(m)^{3/2}} \frac{a_{eq}^{1/2}}{\tilde{r}^{3/2}} (Erf[2 \sqrt{\frac{\pi}{3}} \tilde{r}^{3/2} \sqrt{\mathscr{N}(m)}]- Erf[0])]  
 + \frac{1}{4\pi\mathscr{N}(m)}
\Big]  .
\end{aligned}
\end{equation}    
\end{widetext}
The detectability graphs are obtained by plotting the strain $S_{h}(f)^{1/2}$ (in $Hz^{-1/2}$) versus observed frequency $f_{o} $ (which is $(1+z)^{-1}f_{s} $),  imposing the noise-sensitivity lines of present and future gravitational wave detectors.      
It is important to note  that the mass-density of PBHs in the early Universe very sensitively depends on the quantity $\epsilon $, as given by Eq.(\ref{3.14}). We plot the strain sensitivities ($ S_{h}^{1/2}$) for certain ranges of $\epsilon $ w.r.t. the observed angular frequency in the Figs. \ref{MainDetectability}, \ref{CorrectionTerm1Detectability}, \ref{CorrectionTerm2Detectability}, and \ref{CorrectionTermCrossDetectability} for the four terms, i.e., the main and three correction terms respectively, with the noise sensitivity lines for present and future gravitational wave detectors. The numerical calculations are done with Mathematica (version 9). The numerial  values of the quantities $\mathscr{D} $ and $\alpha $ are estimated by taking the values of $ \Omega_{DE}$ and $\Omega_{M} $ as appoximately 0.68 and 0.31 respectively. These plots are shown below.
\vspace{0.3cm}
\begin{widetext} 
\begin{center}
\begin{figure}
\includegraphics[width=14.5cm]{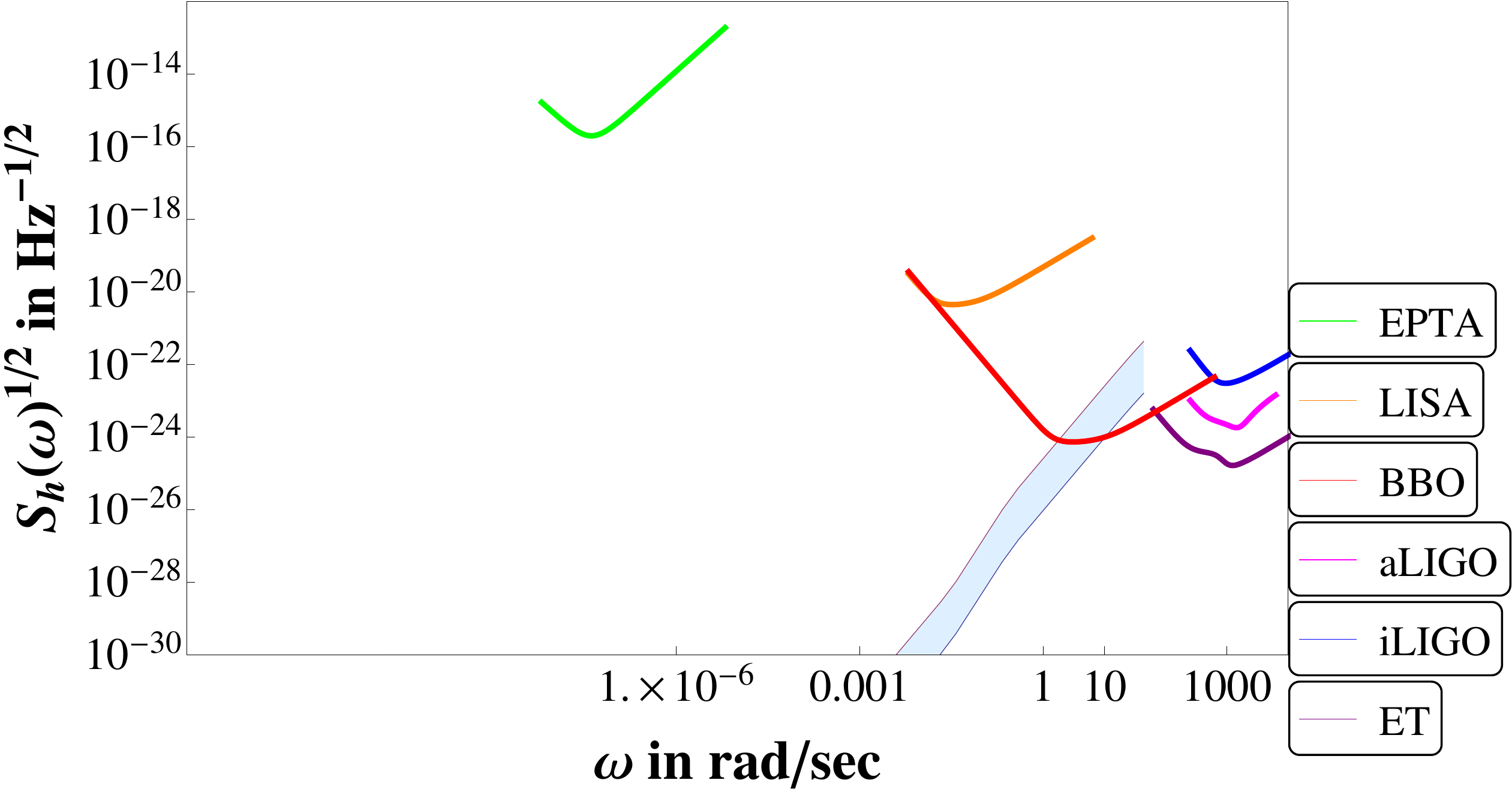}
  \caption{ Plot of the strain $S_{h}(\omega)^{1/2} $ in $Hz^{-1/2}$ of the main term vs the angular frequency (observed) $\omega $ : the band ranges for amplitude of mass-variance of primordial fluctuation $ \epsilon$ from 0.1 to 0.4, for the time t = $10^{-24} \, s $ to 1 s after the big-bang.   }\label{MainDetectability}  
\end{figure}
\begin{figure}
\includegraphics[width=14.5cm]{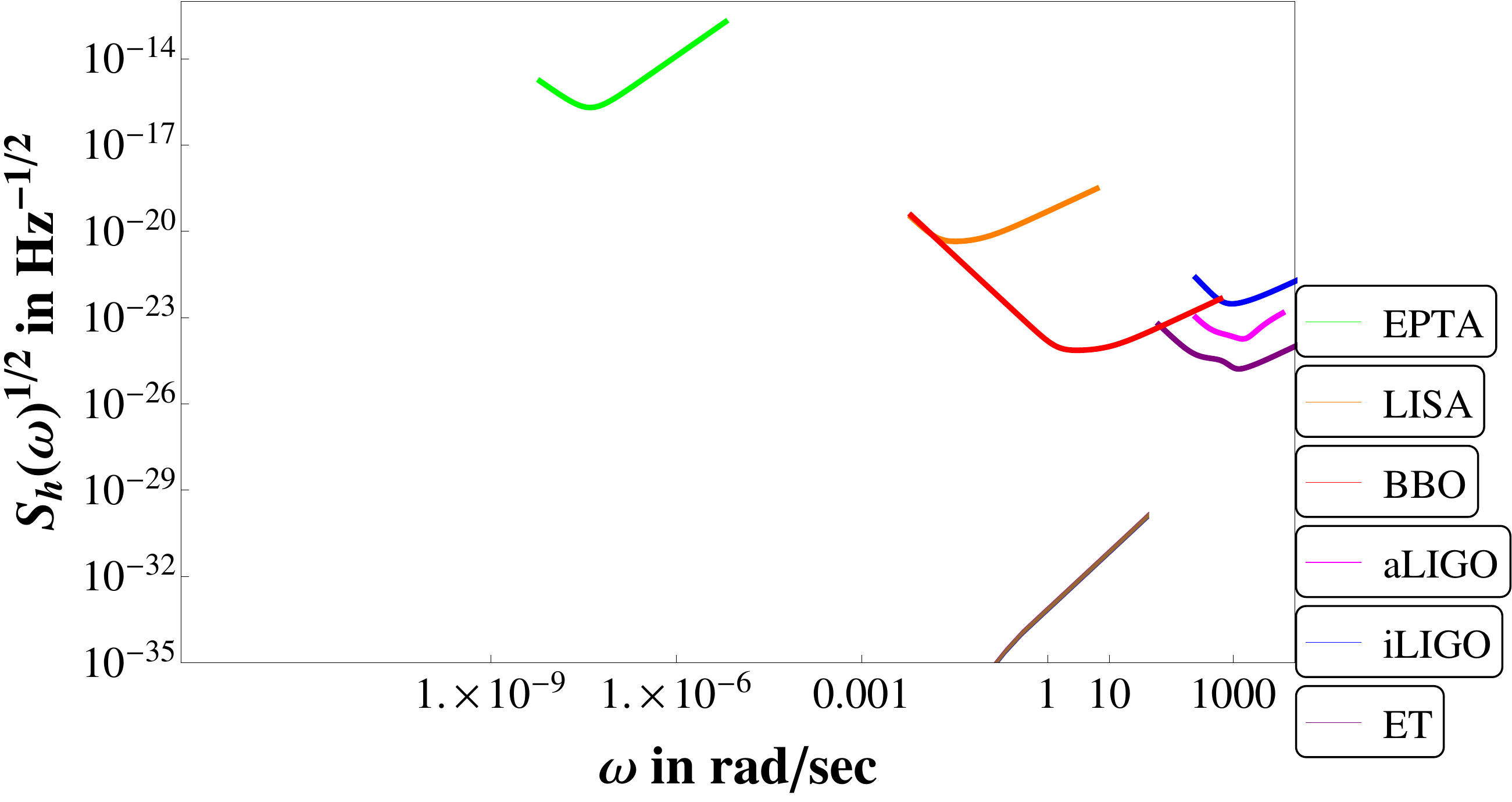}
  \caption{Plot of the strain $S_{h}(\omega)^{1/2} $ in $Hz^{-1/2}$ of the first correction term containing the single time-derivative of the chirp mass vs the angular frequency (observed) $\omega $ : the band ranges for amplitude of mass-variance of primordial fluctuation $ \epsilon$ is 0.4 to 0.8 ; for the time t = $10^{-24} \, s $ to 1 s after the big-bang.   }\label{CorrectionTerm1Detectability}  
\end{figure}
\begin{figure}
\includegraphics[width=14.5cm]{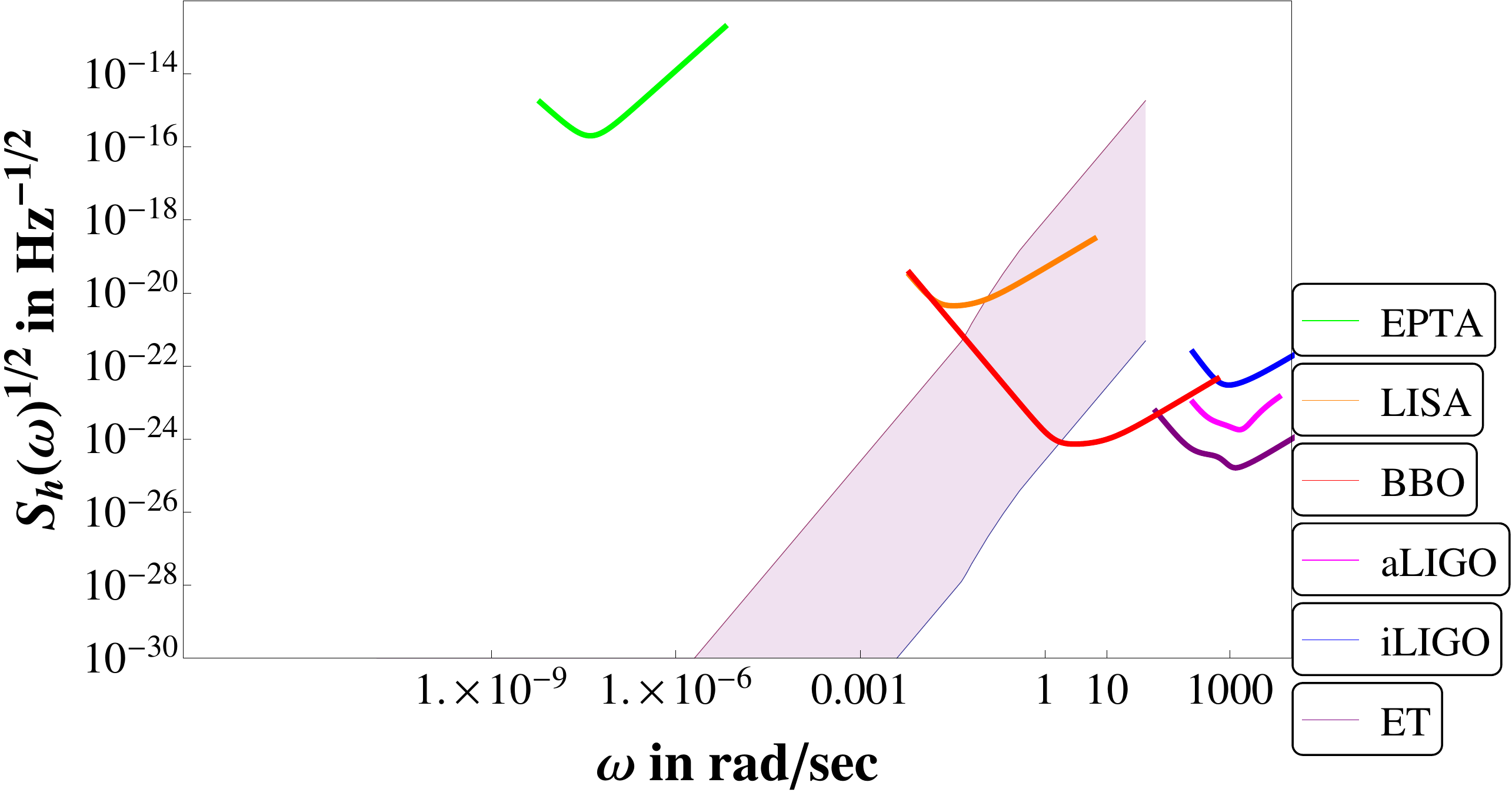}
   \caption{Plot of the strain $S_{h}(\omega)^{1/2} $ in $Hz^{-1/2}$ of the second correction term containing double time-derivative of the chirp mass vs the angular frequency (observed) $\omega $ : the band ranges for amplitude of mass-variance of primordial fluctuation $ \epsilon$ is 0.012 to 0.0125 ; for the time t = $10^{-24} \, s $ to 1 s after the big-bang.   }\label{CorrectionTerm2Detectability}
\end{figure} 
\begin{figure}  
   \includegraphics[width=14.5cm]{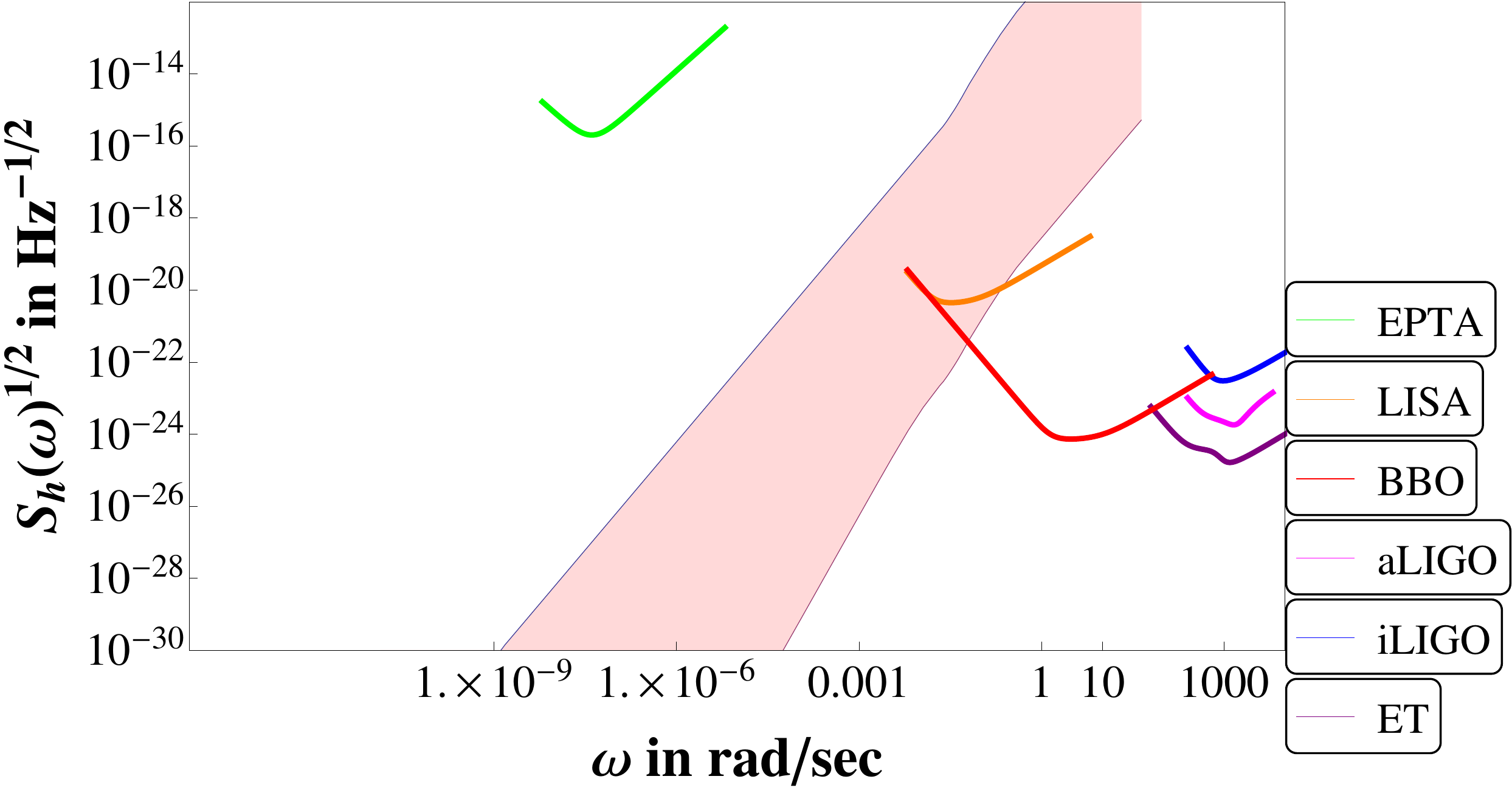}
  \caption{Plot of the strain $S_{h}(\omega)^{1/2} $ in $Hz^{-1/2}$ of the cross-correction term due to non-vanishing product of the second correction term and main term vs the angular frequency (observed) $\omega $ : the band ranges for amplitude of mass-variance of primordial fluctuation $ \epsilon$ is 0.018 to 0.02 ; for the time t = $10^{-24} \, s $ to 1 s after the big-bang.   }\label{CorrectionTermCrossDetectability}   
\end{figure} 
\end{center} 
\end{widetext}  
 \vspace{0.2cm}
 We choose the range of values of the amplitude of mass-variance of primordial fluctuation $\epsilon $, for  the strain sensitivity vs observed angular-frequency band-plot to be such that the strain sensitivity (in $Hz^{-1/2} $) has the value within $10^{-12}$ to $10^{-30}$ $ Hz^{-1/2} $ which is the region where the noise-curves of most of the present and future gravitational wave detectors lie. As we know, that to be detectable, the strain produced by a gravitational wave signal must be above the noise-curve of the associated detector. Only in the case of first correction term, we have extended the lower limit of the strain sensitivity in the detectability graph to $10^{-35} \, Hz^{-1/2} $, because even with very high values of $\epsilon $, we get the strain sensitivity below $10^{-29} \, Hz^{-1/2} $ for the first correction term.
 
 In the fig. \ref{MainDetectability}, the strain sensitivity for the main term of the stochastic background  has been plotted w.r.t. corresponding observed angular frequency  for the range of amplitude of mass-variance of primordial fluctuation $\epsilon $ from 0.1 to 0.4. The noise-curves for different present and future gravitational wave detectors have been shown in the figures \cite{Sathya}. They are iLIGO (initial LIGO), aLIGO (Advanced LIGO), LISA, ET, BBO, and EPTA .
  We  see that certain parts of the stochastic gravitational wave background due to the main term, for the specified range of $ \epsilon$, should be detectable by future gravitational wave detector BBO. 
  
 In fig. \ref{CorrectionTerm1Detectability}, a similar plot of the first correction term is shown. Here, we have shown the band of strain sensitivity vs observed angular-frequency for the range of amplitude of mass-variance of primordial fluctuation $\epsilon $ from 0.4 to 0.8. It can be clearly seen that even for this range of $\epsilon$ with such high values, no region of the stochastic gravitational wave background due to the first correction term is detectable by any present or 
 planned future gravitational wave detector. 
 
 In the fig. \ref{CorrectionTerm2Detectability} and fig. \ref{CorrectionTermCrossDetectability},  similar plots of the second correction term and the cross-correction term  have been shown. In case of the figure \ref{CorrectionTerm2Detectability}, the band of strain sensitivity vs observed angular-frequency has been shown for the range of $\epsilon $ from 0.012 to 0.0125 and in case of figure \ref{CorrectionTermCrossDetectability}, the range of $\epsilon $ is from 0.018 to 0.02. We see that in these 
 cases certain portions of the stochastic gravitational wave background are  detectable by LISA and BBO. 

Note that the range of values of $\epsilon $, chosen for the main term in figure \ref{MainDetectability}, is 0.1 to 0.4, and for the first correction term in figure \ref{CorrectionTerm1Detectability}, is 0.4 to 0.8, which are an order larger than those for the second correction term and cross correction terms, in figures \ref{CorrectionTerm2Detectability} and \ref{CorrectionTermCrossDetectability} respectively (where the ranges are 0.012 to 0.0125 and 0.018 to 0.02 respectively). Yet, we get greater strain for the second correction term and cross correction term than the main term and first correction term. This clearly establishes the dominance of the second and cross correction terms over the main term.

\section{Summary and Conclusions}

In this work we have investigated the stochastic gravitational wave background produced by binaries of primordial
black holes during their early inspiral stage while accreting high density radiation surrounding them  in the early universe. It has been shown that the gravitational wave amplitude  has correction terms because of the rapid rate of increase in masses of the primordial black holes. These correction terms arise  due to non-vanishing first and second time derivatives of the masses and their contribution to the overall double time derivative of the quadrupole-moment tensor. We have found that some of these correction terms are not only significant in comparison with the main term, but even dominant over the main term for certain ranges of time in the early Universe. The significance of these correction terms is not only for the gravitational wave amplitude produced from an individual PBH-binary, but persists for the overall stochastic gravitational wave background produced from them.

We have further studied the detectability of the above stochastic gravitational wave background with present and future gravitational wave detectors. We find that it is possible for such contributions to the overall  stochastic gravitational wave background  to be directly detected with some of the future gravitational wave detectors. Moreover, it would be relevant to study the gravitational wave spectrum emitted from merger stages of such PBH binaries, which should be in the detectability range of aLIGO. Such an occurrence would thus, open up a direct window to probe the early Universe. 

The significant correction terms in the spectral density generated due to rapid increase of masses of the PBHs in the binaries are explicit functions of the density of radiation at the concerned time. Hence, through these correction terms one may be able to constrain the density of radiation at a specific era in early Universe, if the stochastic background is detected in future. 
Moreover, observations of the stochastic background would provide direct clues of  the PBH-mass
ranges, rate of formation of PBH-binaries and their merging-rates, shedding light on the long-standing question as to whether some PBHs still exist in present era of our Universe comprising a fraction of the dark matter. On the other hand, if such a background is not detected, it will help setting  upper limits on the PBH density in early Universe, or more fundamentally, the amplitude of mass variance of the primordial density fluctuations in the  early Universe.

\end{small}
\end{document}